\documentclass{emulateapj}

\newcommand{\mjb}{mJy~beam$^{-1}$}

\shorttitle{Spectral Breaks in Pulsar Wind Nebulae}
\shortauthors{Bock \& Gaensler}

\begin{document}

\title{Measurement of Spectral Breaks in Pulsar Wind Nebulae with
Millimeter-wave Interferometry}

\author{D.\ C.-J.\ Bock\altaffilmark{1}}
\affil{Radio Astronomy Laboratory, University of California, 
  Berkeley, CA 94720}
\email{dbock@astro.berkeley.edu}
\and
\author{B.\ M.\ Gaensler}
\affil{Harvard-Smithsonian Center for Astrophysics, 60 Garden Street
  MS-6, Cambridge, MA 02138}
\email{bgaensler@cfa.harvard.edu}

\altaffiltext{1}{present address: CARMA, P.O Box 968, Big Pine, CA 93513}

\begin{abstract}

  We have observed pulsar wind nebulae in the three supernova remnants
  G11.2$-$0.3, G16.7+0.1, and G29.7$-$0.3 at 89 GHz with the
  Berkeley-Illinois-Maryland Association Array, measuring total flux
  densities of two of them for comparison with archival data at other frequencies. In G16.7+0.1, we find a break in the
  spectrum of the PWN at $\sim 26$ GHz. In G29.7$-$0.3, our data
  suggest a break in the integrated spectrum of the central nebula at
  $\sim 55$ GHz, lower than previously estimated.  However, we have
  found spatial structure in the spectrum of this nebula. The emission
  to the north of pulsar J1846$-$0258 has a broken spectrum, with
  break frequency $\lesssim 100$ GHz, consistent with a conventional
  pulsar-powered nebula. The emission to the south of the pulsar has a
  near-power-law spectrum from radio to X-rays: this component may be
  unrelated to the PWN, or may be evidence of asymmetries and/or time
  evolution in the pulsar's energy output. We present 89 GHz images of each
  remnant.
\end{abstract}


\section{Introduction}

Pulsars are born spinning with up to $10^{51}$~erg of rotational
kinetic energy --- as much energy as in the supernovae in which they
are born. This vast reservoir of energy is ultimately deposited into
the ambient medium through the pulsar's relativistic wind. In many
cases, this interaction between the pulsar and its environment is
directly observable, in the form of a synchrotron-emitting pulsar wind
nebula (PWN), of which the Crab Nebula is the best-known
example. While many pulsars have their beams directed away from us,
PWNe radiate isotropically.  Thus PWNe are powerful probes of pulsars
and their energy loss, even when the pulsar itself cannot be detected.

At radio frequencies, PWNe typically have flat spectra $-0.3 < \alpha
< 0$ ($S_\nu \propto \nu^\alpha$), but in the X-ray band we generally
see $\alpha < -1$. It is therefore commonly presumed that PWNe have at
least one spectral break at intermediate wavelengths.  Such breaks are
expected from theoretical considerations, and result from a
combination of synchrotron losses and the time-evolution of the
pulsar's changing energy output \citep{pac73,reychev84,wol97}.
Locating these breaks gives insight into the physical conditions
of the pulsar wind: in PWNe powered by young ($<5$~kyr) pulsars, the
spectral break can be used to infer directly the nebular magnetic
field strength (e.g.\ \citealt*{man93}), while spatial variations in
the break energy can be used to identify and map out the processes of
particle diffusion and radiative losses within the flow
\citep*{ama00,boc01}.

Most flux density measurements of PWNe to date have been in the X-ray
and low-frequency radio bands. To pin down the break frequencies it is
necessary to measure flux densities in the part of the spectrum near
the spectral breaks, i.e.\ at millimeter and infrared wavelengths.

Here we report on observations of three PWNe for which extant data at
radio and X-ray wavelengths implied a spectral break in or below the
mm band. The PWNe, in the supernova remnants G11.2$-$0.3, G16.7+0.1,
and G29.7$-$0.3 (Kesteven 75), were identified from the catalog of
\citet{gre04}
as being sufficiently bright
and compact to allow imaging with millimeter interferometers.  Two of
the PWNe (in G11.2$-$0.3 and G29.7$-$0.3) contain young X-ray pulsars
\citep{tor97,got00}. Only a few PWNe have been reliably imaged at
millimeter wavelengths with single dish telescopes, and prior to this
study only one (G21.5$-$0.9; \citealt{boc01}) had been imaged with a
millimeter interferometer.

We have been able reliably to image two of the sources observed (in
the supernova remnants G16.7+0.1 and G29.7$-$0.3), obtaining their
total flux densities at 89 GHz. This has allowed us to determine the
frequency of their spectral breaks at radio wavelengths using archival
data and assuming the presence of a solitary break. We have been able
to resolve the PWN in G29.7$-$0.3, letting us compare the
millimeter-wave structure of the PWN with that in the X-ray band and
at lower radio frequencies. We also present a millimeter-wave image of
the SNR G11.2$-$0.3, which has a PWN with radio spectral index
$\alpha=-0.25^{+0.05}_{-0.10}$ \citep*{tam02}.  However, its extension
compared to the spatial sensitivity of the array precludes any
quantitative analysis.

\clearpage

\section{Observations}
\label{sec:obs}

Observations were obtained at 88.6 GHz using the BIMA
Array\footnote{The BIMA Array was operated until June 2004 by the
  Berkeley-Illinois-Maryland Association with support from the
  National Science Foundation.} \citep{wel96} during 2002 and 2003
  (Table~\ref{tab:obs}).  To achieve suitable $uv$ coverage, we used
  the technique of multi-frequency synthesis over both 800 MHz
  sidebands of the local oscillator. We measured only the left
  circular polarization (in the sense of \citealt{ieee69}) in the
  assumption that circularly polarized emission from the sources would
  be negligible.\footnote{The BIMA Array natively measured one linear
  polarization. To convert to circular polarization, 90-GHz
  quarter-wave plates were
  used.}  In order to image fully the extended structure, and to
  assist with the recovery of low spatial frequency information, we
  imaged each source with a 7-point hexagonal mosaic. The primary beam
  of the 6.1-m antennas is Gaussian with FWHM $2\farcm 13$. The short
  interferometer spacings (as low as 6.1 m when antennas are nearly
  shadowed) available from the array provide significant sensitivity
  on spatial scales up to an arcminute. The shortest physical
  baselines are 8.1 m.

The bright quasars QSO B1730$-$130 (for G11.2$-$0.3 and G16.7+0.1) and
QSO B1741-038 (for G29.7$-$0.3) were observed as phase calibrators at
25 to 30 minute intervals, after 3 complete cycles of 7 pointings. We
expect the astrometric accuracy of the images to be better than 1
arcsec, a precision routinely obtained with the BIMA Array
\citep{loo00}.  Single sideband system temperatures during the
observations were between 200 and 600 K (scaled to outside the
atmosphere). We used the on-line absolute flux density scale
determined from many observations of planets, which have been found on
average to be stable over time \citep{muh91}. As a check we
observed Uranus for 10 minutes during each observing session, finding
a day-to-day scatter for the mean gain on all antennas between 0.95
and 1.12, implying multiplicative corrections to the flux density
scale of between 0.91 and 1.25. The mean correction would be 1.09 for
observations of G16.7+0.1 and 1.05 for G29.7$-$0.3. The gains for each
sideband were consistent within a few percent on each day. In this
paper we retain the on-line scale, and note that with several days'
observations contributing to each image our measurements are
consistent with an uncertainty in the overall flux density scale of
about 10\%.

Imaging was done with the MIRIAD software package (\citealt*{sau95}),
using SDI deconvolution \citep*{ste84}: such CLEAN-based
algorithms are generally superior to maximum entropy methods for low
signal to noise ratios \citep*{hel03}. The multiple pointing centers
were combined into a mosaiced image corrected for the primary beam. To
assess the reliability of flux density recovery when imaging these
sources, simulations of each observation were made with MIRIAD, using
the actual observations as a template. The results of the simulations
are presented with the data on each source in the next section.

\begin{deluxetable*}{lclcccc}
\tabletypesize{\footnotesize}
\tablewidth{\textwidth}
\tablecaption{Summary of observations\label{tab:obs}}
\tablehead{
& &  & Time on && \multicolumn{2}{c}{Synthesized Beam} \\
\cline{6-7}
&  Antenna & & Source &Image rms & FWHM size   & PA  \\
SNR & Config. & Date(s) & (hr) & (\mjb)  & (\arcsec)  &  (\degr) 
}
\startdata
G11.2$-$0.3  & D       & 2002 Jun 10                       & 3.7
&$\la 10 \tablenotemark{a}\phm{\la}$
                                                                                            & 18.9$\times$17.6     & 19.0 \\
G16.7+0.1    & D       & 2002 Jun 2, Sep 10; 2003 Sep 1,4  & 14.3\phm{1} & 4       & 25.2$\times$18.7     & \phn 9.7\\
G29.7$-$0.3  & D       & 2002 May 30, 2003 Aug 31          & 8.2                  & \phm{.7}1.7\tablenotemark{b}      & 13.3$\times$10.4\tablenotemark{b} 
                                                                                                                   & 17.1\tablenotemark{b}\\
             & C       & 2002 May 9,16                     & 7.9                  &         &                      & \\
             & B       & 2002 Mar 15                       & 5.6                  &         &                      & \\             
\enddata
\tablenotetext{a}{\phm{}Low-level emission throughout the field makes the rms difficult to determine accurately}
\tablenotetext{b}{For C and D configuration observations}
\end{deluxetable*}

\section{Results}

\subsection{G11.2$-$0.3}

\begin{figure}
\centering\includegraphics[width=3.375in]{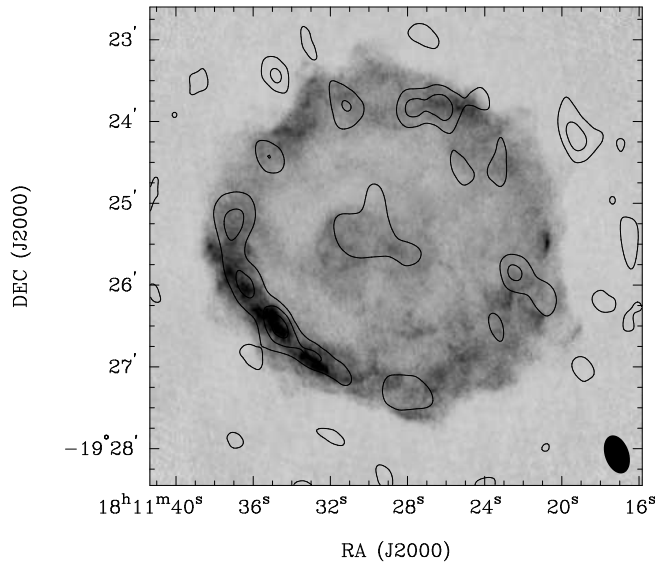}
\caption{Image of G11.2$-$0.3 at 89 GHz (contours) and 8.6 GHz
  (grayscale; \citealt{rob03}). The contours are at 10, 20, and 30
  \mjb. The synthesized beam at 89 GHz is shown at the lower right.}
\label{fig:g11}
\end{figure}

The BIMA image of G11.2$-$0.3 is shown with contours in
Figure~\ref{fig:g11}, overlaid upon an 8.6 GHz image of
\citet{rob03}. Based on positional coincidence with the 8.6 GHz
emission, we appear to have detected the PWN in this remnant (the
single contour at the center of the image). However, the extended
structure in the PWN is not well imaged by the array: the modeling
described in section~\ref{sec:obs} showed that we have measured
only approximately 10\% of the emission, indicating that the PWN is
nearly resolved out on all baselines. Given the magnitude and
unreliability of the correction factor required to estimate the true
flux density, we do not believe that these data can provide a reliable
estimate of the flux density of the PWN. High sensitivity single dish
measurements would be required to investigate this object further.

\subsection{G16.7+0.1}

\begin{figure}
\centering\includegraphics[width=3.375in]{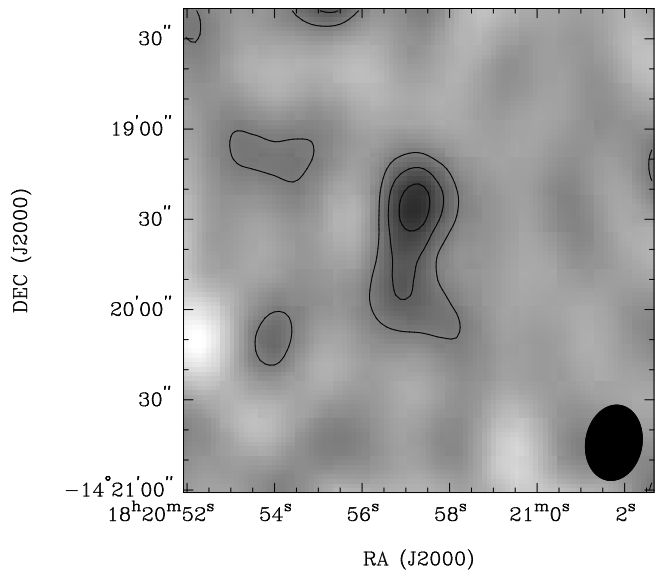}
\caption{Image of G16.7+0.1 at 89 GHz. The contours are at 4, 
  8, and 12 \mjb. The
  synthesized beam at 89 GHz is shown at the lower right.}
\label{fig:g16}
\end{figure}

The PWN in this remnant (Figure \ref{fig:g16}) is detected as a peak of 14 \mjb\ at 8$^{\rm h}$20$^{\rm m}$57\fs 2
$-$14\degr 19$'34''$ (J2000).  The feature is extended: an elliptical
Gaussian fitted to the feature has a size of $51''\times 20''$ (at
position angle 6\degr\ east of north), and an integrated flux density
of 27 mJy. The size and position are consistent with the measurements
of \citet{hel89}.

To assess the reliability of this measurement we have modeled the
response of the array to two simple sources: a $25''\times 7''$
Gaussian with major axis 10\degr\ west of north and a $45''\times
15''$ Gaussian with major axis 20\degr\ west of north, corresponding
approximately to the 50\% and 20\% contours in the 2 cm image of
\citet{hel89}. In deconvolved images we recovered 97\% and 80\% of the
total emission respectively, without taking into account a small
residual negative background. We conclude that we should have measured
at least 90\% of the emission from this object, and consider that we
have reliably determined the flux density of the PWN at this
frequency. Taking into account the uncertainty in the flux density
scale (\S \ref{sec:obs}) and the low signal to noise we estimate the
total uncertainty in our measurement to be no more than 20\%. 

\subsection{G29.7$-$0.3 (Kesteven 75)}
\label{sec:results_g29}

\begin{figure}
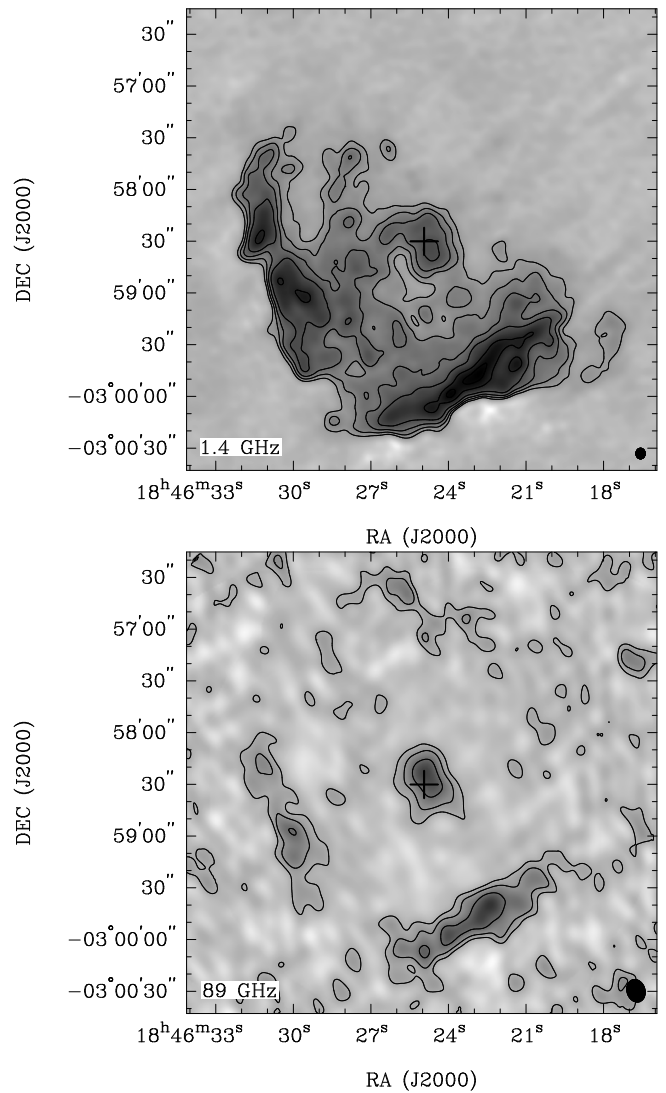

\centering
\includegraphics[width=3.375in]{f3a.eps}
\includegraphics[width=3.375in]{f3b.eps}
\caption{Images of G29.7$-$0.3 at 1.4 GHz (\citealt{hel03g29};
  contours  2, 5, 10, 20,
  50, 80 \mjb\ on logarithmic grayscale) and 89 GHz (linear scale;
  contours at 2, 5, 10, 20 \mjb). The position of the pulsar
  J1846$-$0258 \citep{got00} is marked with a cross. The synthesized 
  beams are shown at the lower right of each image.}
\label{fig:g29}
\end{figure}

The BIMA image of G29.7$-$0.3 is presented in Figure \ref{fig:g29}.
For comparison, the 1.4 GHz image of \citet*{hel03g29} is shown also.
This object is sufficiently bright at millimeter wavelengths that we
have been able to image it using both the BIMA C and D configurations.
However, the data obtained with the B configuration did not add
significantly to the results: they have not been included in the image
in Figure \ref{fig:g29}, nor in the quantitative analysis that follows.
In the figure, both the PWN (at the image center) and the partial shell
(to the south and east) can easily be seen at each frequency.  As
expected, the PWN is relatively brighter than the shell at the higher
frequency, i.e.\ it has a flatter spectral index. Note that the PWN is
central in the corresponding X-ray image \citep{hel03g29}, which also
shows the northern extent of the SNR's shell.

We have measured an 89 GHz flux density for the PWN of 80 mJy in 0.6
arcmin$^2$ (within approximately the ``0 \mjb\ contour''), taking into
account a local background (a ``negative bowl'') within the SNR shell
of about $-$36 mJy~arcmin$^{-2}$. To quantify the degree to which the
array is sensitive to the extended emission, we simulated an
observation of the PWN using the 1.4 GHz image \citep{hel03g29} as a
model.\footnote{We chose this image for a model owing to its higher
  signal to noise.  Recalling the flat PWN spectrum in the radio, we
  do not believe that the difference between the
  PWN shape at the two frequencies will significantly affect our
  result.}  We added Gaussian noise to the undeconvolved image to give
  a similar signal to noise ratio. After deconvolution, we recovered
  92.4\% of the flux density in the simulated observation. Applying
  the correction for interferometric filtering of the extended
  emission thus determined, we obtain a flux density for the PWN of 86
  mJy.  In the simulation the negative background was much less
  pronounced than in the BIMA observation: this artifact is thus
  probably due primarily to poor deconvolution of emission from the
  SNR shell, which was not included in our simulation.  We consider an
  upper limit to the imaging errors to be 10\%; combining this with
  the 10\% flux calibration uncertainty (\S\ref{sec:obs}) leads to a total uncertainty
  of 15\%.

At the nearby frequency of 84 GHz, the observations of \citet{sal89b}
imply an integrated flux density of 159 mJy, inconsistent with the
present data. However, the earlier observations were made using a
single dish with a beam of 76$'' \times 70''$, substantially larger
than the characteristic source size ($\sim 30''\times
20''$). \citeauthor{sal89b} corrected their measurement for the
smaller source size to derive their integrated flux density. However,
it seems possible that Salter et al.\ also measured some shell emission.
The advantage of the present interferometric measurement is that it
would have filtered out any contaminating emission from the SNR's
shell, and should provide a more reliable determination of the flux density
attributable to the PWN alone.

\section{Discussion}

\subsection{G16.7+0.1}

\begin{figure}
\centering
\includegraphics[width=3.375in,clip=]{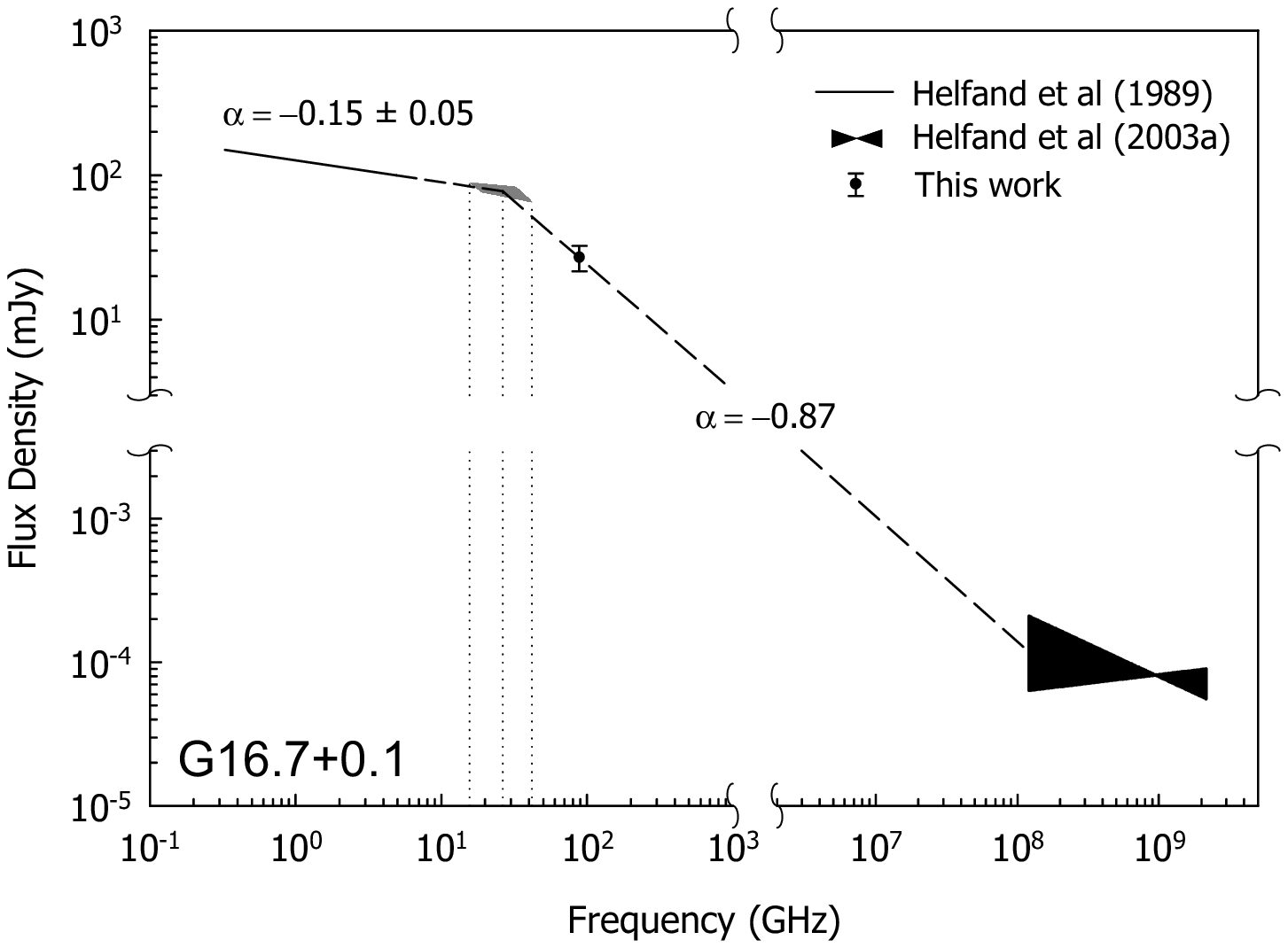}\\
\vspace*{0.2cm}
\includegraphics[width=3.445in,clip=]{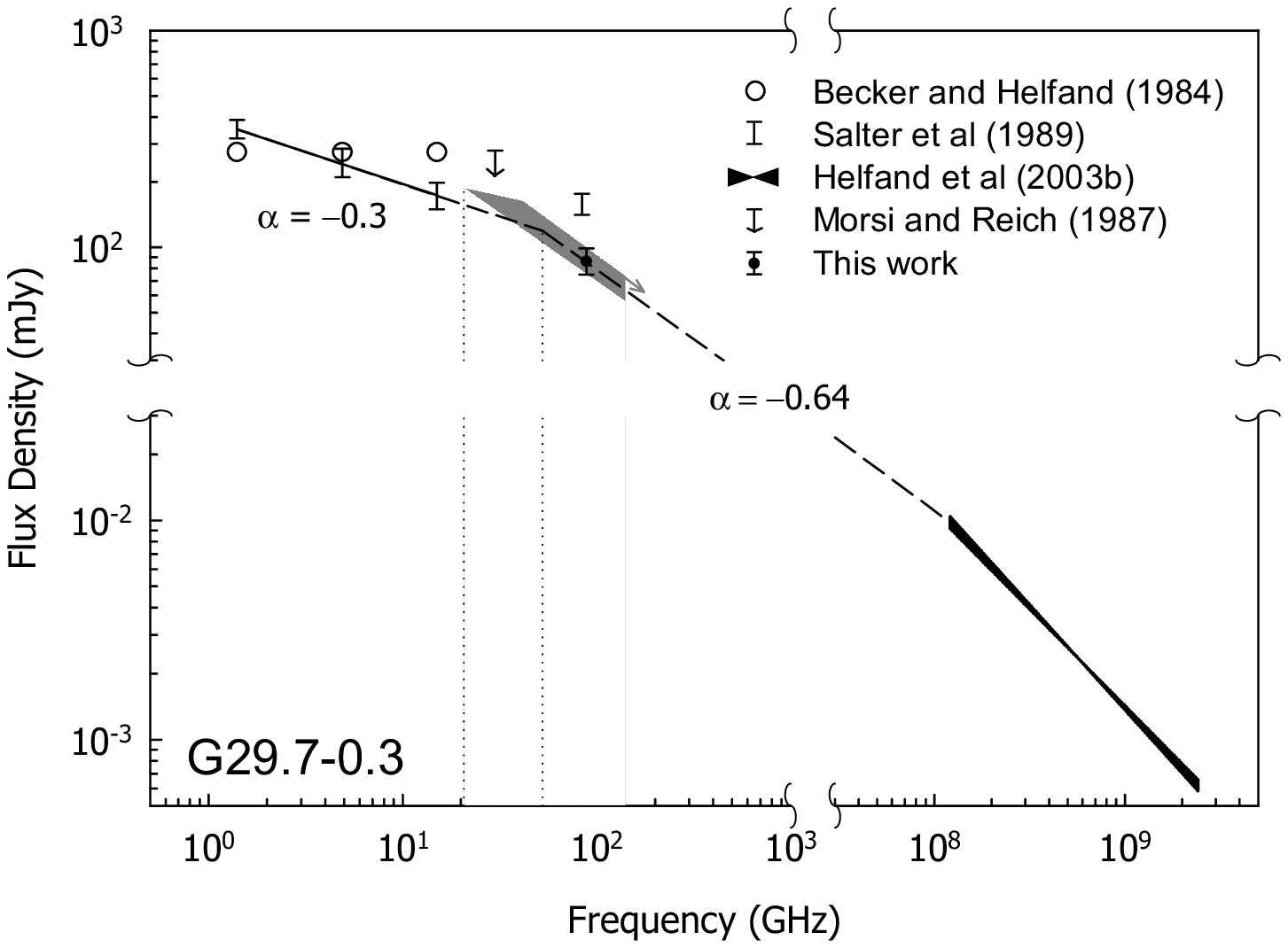}
\caption{Integrated spectra of the pulsar wind nebulae in G16.7+0.1 and
  G29.7$-$0.3. In regions where the spectra are well determined they
  are shown with unbroken lines. Elsewhere simple straight line
  spectra implied by sparse measurements are shown by a dashed line.
  Uncertainties are shown where available; at X-ray wavelengths these
  are indicated by black shading. The regions within which spectral
  breaks are implied are shown by gray shading, and bounded by the
  vertical dotted lines (see text for details).}
\label{fig:spectra}
\end{figure}

A spectrum of the PWN in G16.7+0.1 is shown in
Figure~\ref{fig:spectra}.  There are few measurements available of the
flux density of this PWN. The solid line at low frequencies represents
the radio spectrum derived from observations at 6, 20, and 90 cm by
\citet{hel89}. Recent X-ray measurements \citep*{hel03g16} imply that
there must be at least one break between the radio and X-ray regimes;
the present measurement implies that this is at about 26 GHz (central
dotted line). In the figure, the gray shaded area represents the region
of intersection of the Helfand et al.\ radio spectrum and a simple
radio/X-ray spectrum (i.e. assuming a single break between X-ray and radio wavelength) consistent with the 89 GHz and X-ray
uncertainties. This provides one way to quantify the uncertainty in
the break frequency: the region has minimum and maximum frequencies
(shown by the outer dotted lines) of 16 and 42 GHz respectively,
leading to an estimate of the break frequency of $26^{+16}_{-10}$ GHz.

The PWN in G16.7+0.1 has one of the lowest known break frequencies in
a PWN: breaks at lower or comparable frequencies have been reported
only in G27.8+0.6 (between 5 and 10 GHz; \citealt*{rei84}) 
and G74.9+1.2 (between 10 and 30 GHz; \citealt{mor87a}).
Surprisingly, the spectrum within the X-ray band ($\alpha=-0.17\pm
0.29$) is flatter than the radio to X-ray spectrum. Presumably this is
not the result of inverse Compton scattering, which is generally
expected to be significant at $\gamma$-ray wavelengths. This
phenomenon has been seen in one other PWN, that around the pulsar PSR
B1757$-$24 \citep{kas01}.\footnote{A step like that in the spectrum of
  the Crab Nebula at infrared/optical wavelengths, thought due to dust
  reradiation or extinction \citep{man77}, would not show up in the
  bulk radio to X-ray spectral index in the same way.} In that case,
the X-ray and radio spectra are also approximately flat, with a
steeper spectrum required to connect the radio and X-ray observations.
However, whereas \citeauthor{kas01}\ found that the X-ray spectrum was
just marginally consistent with the implied radio to X-ray average
spectrum, this possibility is very unlikely in the case of G16.7+0.1,
where the average radio to X-ray spectrum is not consistent with the
X-ray data even within the 90\% uncertainty limits of
\citet{hel03g16}.

One explanation for spectra of this shape is given by
\citet{reychev84}. They calculate that a combination of older
particles suffering synchrotron losses due to compression of the
magnetic field by the reverse shock and particles injected after the
shock can give rise to a spectrum of this shape.  However, whereas the
reserve shock is expected to arrive after about $10^4$ years,
\citet{hel03g16} have estimated G16.7+0.1 to be only about 2000 years
old. Therefore this seems like an unlikely explanation for this
source.

\subsection{G29.7$-$0.3: Integrated Spectrum of the PWN}

An integrated spectrum of the central nebula in G29.7$-$0.3 is plotted
in Figure \ref{fig:spectra}. This object was first seen by
\citet{bec76}, who measured a radio spectral index of $-0.27$, flatter
than that of the SNR shell. VLA observations \citep{bec84} confirmed
the discovery but implied a flat ($\alpha=0.0$) spectrum. Hunt et al.\
(as reported by \citealt{sal89b}) from different VLA observations
obtained integrated flux density measurements that imply a radio
spectral index of $-0.30$. Both sets of VLA observations are
plotted in Figure \ref{fig:spectra}. We have measured a flux density
of 343 mJy for the PWN in the recent 1.4 GHz data of \citet{hel03g29},
consistent with the Salter et al.\ result. 

At higher frequencies, \citet{mor87b} measured 0.28 Jy from the PWN at
30 GHz, but note that they could have suffered from confusion (their
observations had resolution $26\farcs 5$), leading to an
overestimated result. It seems possible that they measured some of
the broad underlying shell-related emission as well, which was not
imaged at 89 GHz, and could have been poorly imaged at lower radio
frequencies. This could have led to a further overestimate. As noted
earlier (\S \ref{sec:results_g29}) we consider the \citeauthor{sal89b}\ 84 GHz
measurement also to be an overestimate, for similar reasons.

Combined with the Salter et al.\ lower-frequency measurements, the
X-ray data of \citet{hel03g29}, and assuming a simple unbroken
spectrum between radio and X-ray wavelengths, our observations imply a
spectral break at approximately 55 GHz (shown by the right-hand
vertical dotted line in the figure). Assuming the break to be sharp,
the integrated flux density of the PWN at this frequency would be
about 120 \mjb.  Allowing the spectrum above and below the putative
break to vary within the uncertainties of the measurements implies
that the break should fall within the shaded gray area, with a minimum
frequency of 21 GHz (left-hand vertical line). The break is not
strictly constrained at high frequencies, since a straight line is
just consistent with the low-frequency Hunt et al.\ measurements and
our 89 GHz measurement.  However, it is likely that the break is below
89 GHz.

We could alternatively have used the centimeter-wave data of
\citet{bec84} and obtained a somewhat lower break
frequency. However, given the agreement between the data of Salter et
al.\ and \citet{hel03g29}, and the fact that the Salter et al.\
data were quoted with uncertainties, we prefer the later measurements.


\subsection{G29.7$-$0.3: Lobes of the PWN?}

\begin{figure*}
\centering
\includegraphics[width=2in,bb=196 291 414 503,clip=]{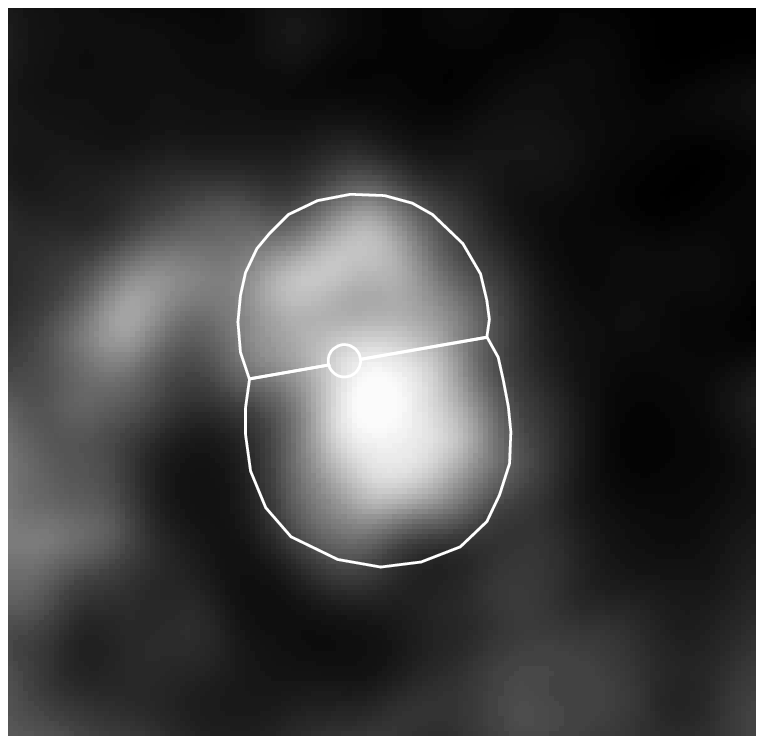}
\includegraphics[width=2in,bb=196 291 414 503,clip=]{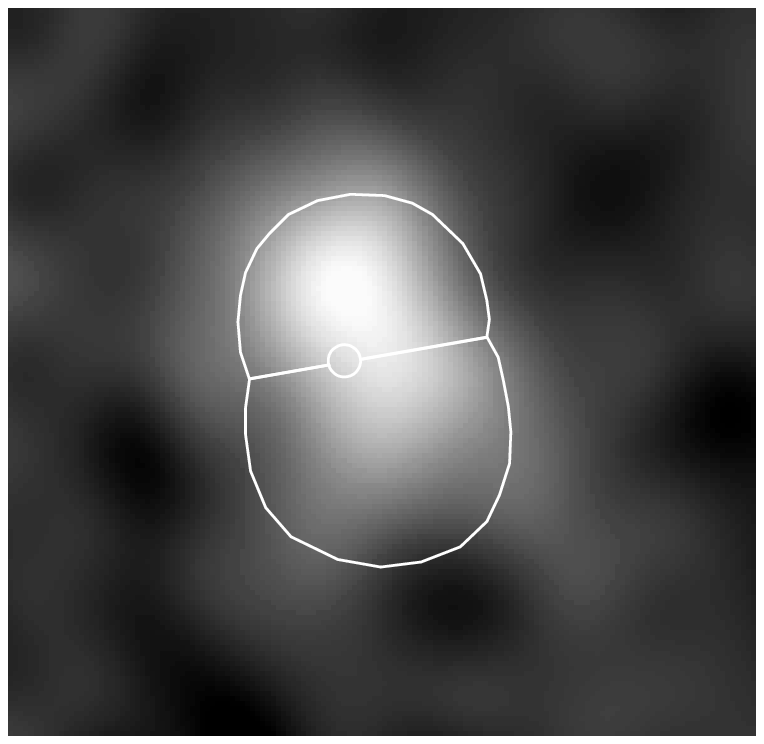}
\includegraphics[width=2in,bb=196 291 414 503,clip=]{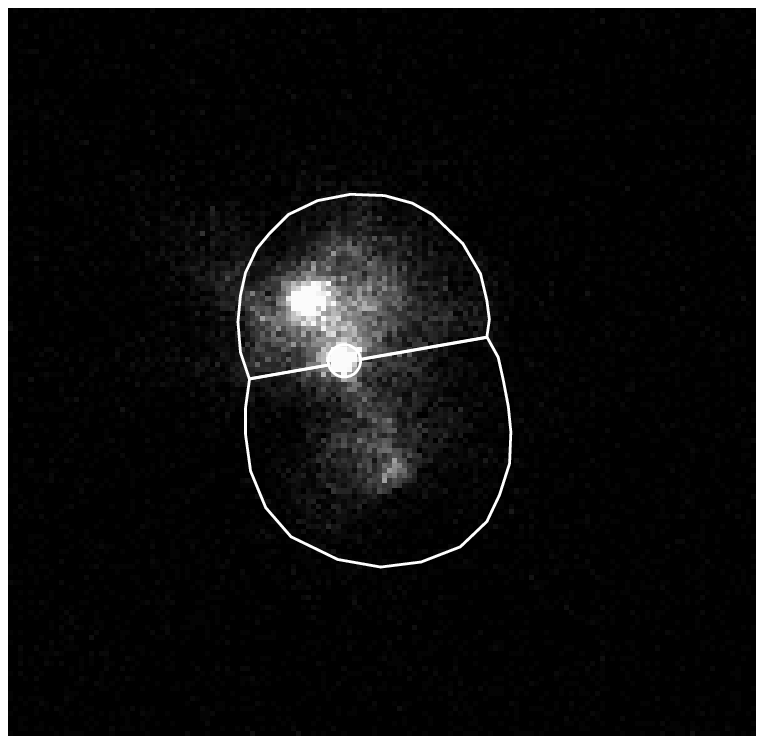}
\caption{Images of G29.7$-$0.3 at (left to right) 1.4 GHz \citep{hel03g29}, 89 GHz, and
  0.5--10 keV (\emph{Chandra} archive; as presented by
  \citealt{hel03g29}), showing the regions over which emission was
  integrated in the southern and northern lobes.}
\label{fig:lobeimages}
\end{figure*}

The high resolution X-ray data clearly show two lobes of emission, to
the north and south of the pulsar. The northern lobe is brighter at
X-ray wavelengths. A comparison of images of the PWN at 1.4 GHz, 89
GHz and in X-rays (Figure \ref{fig:lobeimages}) reveals that the PWN
spectrum is also asymmetric. At 1.4 GHz, the PWN is brighter south of
the pulsar, while at 89 GHz and above it is brighter north of the pulsar.
These two components may have different origins, causing an integrated
spectrum to be difficult to interpret.

In Table~\ref{tab:lobefluxes}, we show flux densities of these
northern and southern lobes, integrated over the regions shown in
Figure~\ref{fig:lobeimages}. The regions chosen were a compromise
between the competing desires to minimize contamination from the
background and to include the majority of emission at each frequency.
The area around the pulsar was excluded from all datasets, avoiding
the need to estimate an X-ray background for normalizing the
integrated areas.  At 89 GHz the same background as in \S\ref{sec:obs}
was subtracted.  \citet{hel03g29} were not able to find a significant
difference between the X-ray spectra of the northern and southern
lobes, so the X-ray flux densities quoted correspond to the luminosity
reported by \citeauthor{hel03g29}, apportioned according to the total
counts in the two regions, assuming a uniform X-ray spectrum across
the PWN.  The 89 GHz uncertainties quoted are as in
section~\ref{sec:obs}, those for X-rays are the standard errors for a
Poisson distribution, and we assign an uncertainty of 10\% to the
1.4 GHz measurements, to account for imaging errors. This is probably
an overestimate of the true uncertainty at 1.4 GHz.

\begin{figure}
\centering
\includegraphics[width=3.375in,clip=]{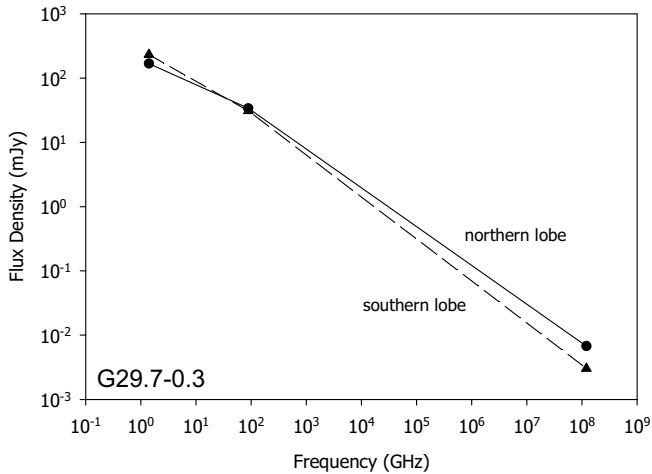}
\caption{Simple ``straight-line'' spectra of the northern and southern 
lobes of the PWN in
  G29.7$-$0.3, derived from the data in Table~\ref{tab:lobefluxes}.  }
\label{fig:lobespectra}
\end{figure}

Given the differing resolutions and systematic errors of the three
datasets\footnote{In the 89 GHz image, there are only 3 beam areas
within the northern region.}, the results are necessarily
qualitative. However, as may be seen in Figure~\ref{fig:lobespectra},
the northern lobe not only has a flatter spectrum on average, but
shows strong evidence for break. Meanwhile, the spectrum of the
southern lobe may be unbroken. In the figure, the uncertainties at
radio frequencies are of order the point size, while those of the
X-ray values are much smaller. The excellent positional agreement of
the SNR shell at 1.4 and 89 GHz confirms the registration of the
images.  We believe that the difference in spectral shape between the
lobes is reliable.

\begin{deluxetable*}{ccc}
\tablewidth{11.5cm}
\tablecaption{Flux densities of the lobes around the pulsar in G29.7$-$0.3\label{tab:lobefluxes}}
\tablehead{
         & \multicolumn{2}{c}{Flux Density (uncertainty)} \\
Frequency& \multicolumn{2}{c}{(mJy)} \\
\cline{2-3}
(GHz)    &  N.\ lobe & S.\ lobe
}
\startdata
\phm{0}1.4               & $170(35)\phm{.00000}$     & $230(35)\phm{.00000}$ \\
89\phn\phd               & $\phn 34(5)\phm{.00000}$ & $\phn 31(5)\phm{.00000}$ \\
$1.2\times 10^8$ (0.5 keV) & $\phm{00}0.00675(5)$ & $\phm{00}0.00303(3)$ \\
\enddata
\tablecomments{Datasets for the measurements at 1.4 GHz and 0.5 keV
  are as referenced in Figure~\ref{fig:lobeimages}}
\end{deluxetable*}

\begin{figure*}
\centering
\includegraphics[width=2in,bb=196 291 414 503,clip=]{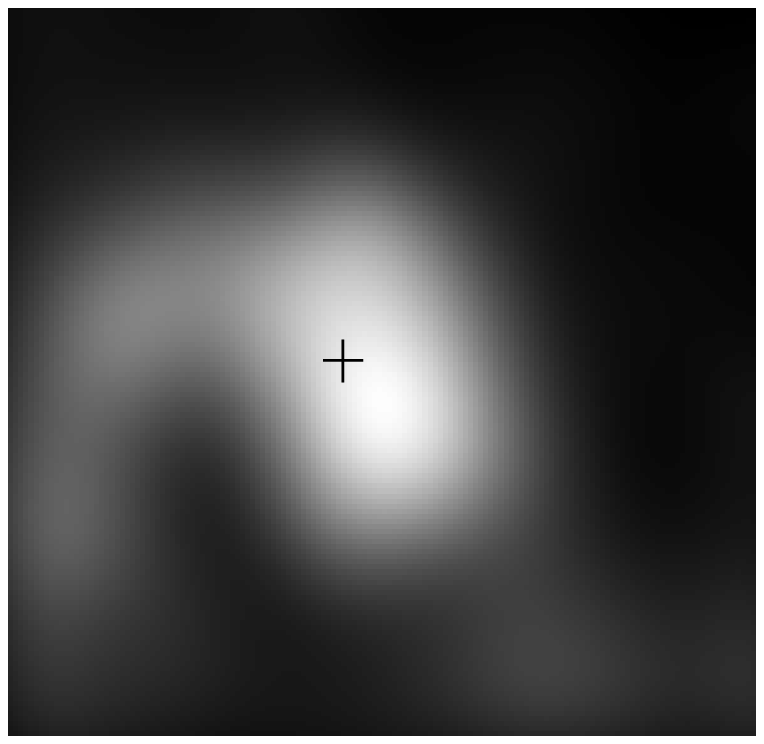}
\includegraphics[width=2in,bb=196 291 414 503,clip=]{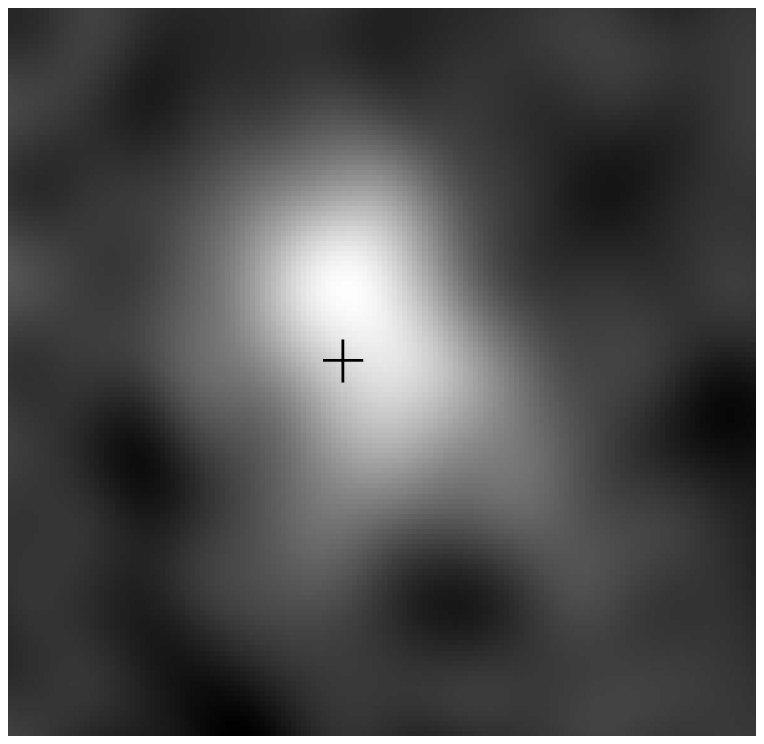}
\includegraphics[width=2in,bb=196 291 414 503,clip=]{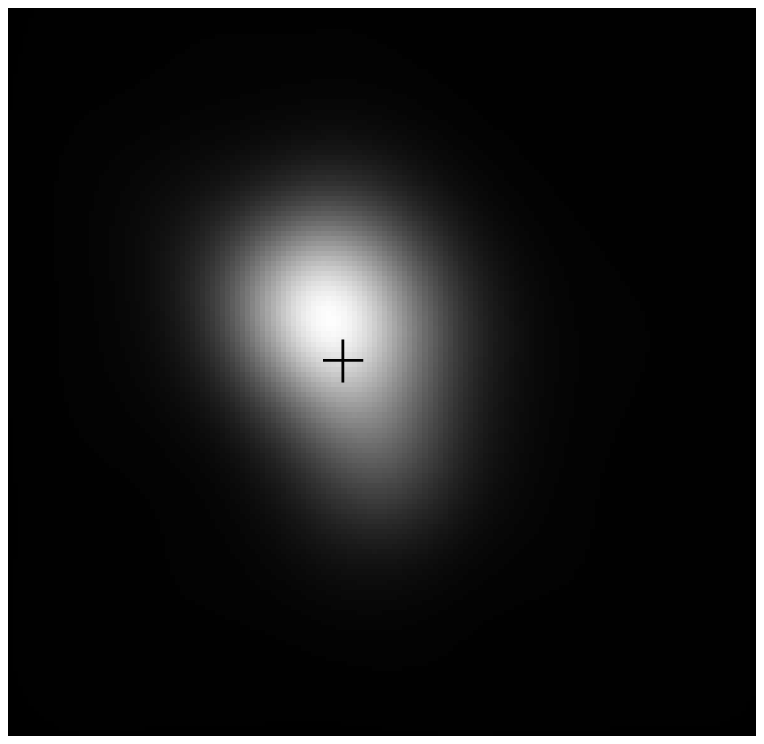}
\caption{As for Figure ~\ref{fig:lobeimages}, all convolved to the
  resolution of the 89 GHz image. The position of PSR J$1846-0258$ has
  been marked with a cross.}
\label{fig:lobeimages2}
\end{figure*}

The same result may be obtained qualitatively by convolving the 1.4
GHz and X-ray observations to the resolution of the 89 GHz data
(Figure~\ref{fig:lobeimages2}). For this analysis, a background
estimated at 24 counts/pixel was substituted for X-ray emission from
the pulsar before convolution. From the figure, it is clear that the
trend of steeper emission to the south holds. Although the systematic
uncertainties are different, this method again implies a break in the
spectrum of the northern lobe, and allows the southern lobe to have a
simple power law spectrum from radio to X-rays.

The simplest explanation for the difference in spectrum to the north
and south of the pulsar is that we are seeing both PWN and shell
emission superimposed. In this scenario, a majority of the
steep-spectrum emission seen at 1.4 GHz is attributed to the shell,
seen near the pulsar due to projection. Other 1.4-GHz emission presumably from
the shell may be seen at to the east of the PWN in
Figure~\ref{fig:g29}. The extent of the true PWN is then seen most
easily in X-rays, where the southern portion is much fainter. 

The asymmetry in the position of the pulsar with respect to the
northern lobe (the implied true PWN) may be due to the passage of a
reverse shock from the interaction of the supernova with the
surrounding medium. An asymmetric medium will cause an asymmetric
reverse shock. This phenomenon has been used to account for the PWN in
the Vela supernova remnant, which is largely to one side of the
presumed progenitor position \citep*{blo01}.  However, the expected
$10^4$ years until the reverse shock is much larger than the age
implied by the pulsar's rate of spin down, which falls in the range
$\sim 700$--1700 years \citep{got00,mer02}. Thus in the absence of an
extremely dense ISM the passage of a reverse shock does not seem a
likely explanation for the asymmetry.

An alternative possibility is that the pulsar may have traveled from
the apparent center of the PWN since the explosion. Assuming that the
pulsar was born at the brightest part of the northern lobe, the
kinematic distance of 21 kpc \citep{bec84} and age range above imply a
transverse velocity of 400--1000 km s$^{-1}$. The upper
end of this range is higher than the transverse velocity of almost all
known pulsars, so improved pulsar timing (and eventually proper motion
measurements) should allow us to evaluate this scenario in the future, although we note that this pulsar has so far been detected only in X-rays.

A final possibility is that the asymmetry in the nebula is intrinsic
to the pulsar wind. In this case the southern lobe could be
unassociated shell emission, as suggested above, or a part of the PWN
with a different spectral shape.

\subsection{Origin of the Spectral Breaks}

In both the PWNe in G16.7+0.1 and G29.7$-$0.3, there is strong
evidence for a low frequency spectral break. For G29.7$-$0.3, this
break may, as mentioned above, occur only in the northern lobe of the
PWN. Both remnants are relatively youthful.  The conventional spectral
break due to synchrotron losses in a field $B$ mG occurs at age $\sim
40B^{-1.5}\nu^{-0.5}$ kyr at frequency $\nu$ GHz. Thus for G16.7+0.1
we expect a field of $\sim 2.5$ mG, while for G29.7$-$0.3 we expect $>
1.5$ mG. These values are substantially higher than inferred in other
remnants from equipartition arguments, or more directly in the Crab
Nebula (e.g.\ \citealt{hes96}),
and seem to indicate that standard synchrotron losses alone are not
the origin of the breaks.

We are left with the conclusion that perhaps these are indeed remnants
of a ``second kind'' \citep{wol97}, in which intrinsic processes
result in a spectrum different from that of the Crab. One possibility
is that the pulsar's excitation has decreased substantially at some
point \citep{gre92}, with the break moving to lower frequencies with
time \citep{wol97}. Alternatively the pulsar injection spectra may not
be a simple power law.

\section{Conclusions}

We have carried out observations of three pulsar wind nebulae at
millimeter wavelengths with the Berkeley-Illinois-Maryland Association
Array. One of these PWNe, in the SNR G11.2$-$0.3, was not adequately
imaged to allow quantitative analysis. We find the break frequency of
the PWN in G16.7+0.1 to be at about $26$ GHz, and see an unusual
upturn in the spectrum below or in the X-ray band. In only one other
PWN, that surrounding PSR B1757$-$24, has a similar effect been
implied.

The bulk spectrum of the PWN in G29.7$-$0.3
indicates a likely spectral break at 55 GHz, but our observations
and previous data reveal a marked variation in spectrum across the
PWN. The spectrum is steeper to the south of the pulsar. It may be
that the southern emission does not represent the continuing effect of
the pulsar, but corresponds to a part of the SNR shell seen in
projection. However, we cannot account for the one-sided PWN under
this scenario.

The number of PWNe for which an analysis like this may be made is
limited by the large size of the PWNe compared to the spatial scales
on which millimeter interferometers are sensitive. However, the
sensitivity and resolution of next generation instruments, such as
ALMA, will allow these investigations to be conducted on much larger
samples in other galaxies. Meanwhile, improved single-dish mapping
techniques and sensitive wide-field imaging interferometers such as
CARMA could enhance the high-frequency data on Galactic PWNe.

\acknowledgements

We thank E.~Gotthelf, D.~Helfand, and M.~Roberts for making available
their data in electronic form, and M.~Wright for useful discussions.
This work has made use of the NASA ADS and was partially supported by
NSF grants AST-9981308 and AST-0228963 to the University of
California, Berkeley.

{}

\end{document}